\documentstyle[12pt]{article}
\def\be{\begin{equation}}
\def\ee{\end{equation}}
\def\ba{\begin{eqnarray}}
\def\ea{\end{eqnarray}}
\def\bq{\begin{quote}}
\def\eq{\end{quote}}

\def\NP{{\it Nucl. Phys.} }

\begin{document}

\renewcommand{\theequation}{\arabic{section}.\arabic{equation}}

\thispagestyle{empty}
\begin{flushright}
UMN-TH-1602/97\\
WATPHYS-THY-97/06\\
hep-th/9708008\\
August 1997
\end{flushright}
\vspace*{1cm}
\begin{center}
{\Large \bf  Singularities In Scalar-Tensor Cosmologies}
 \\
\vspace*{1cm}
Nemanja Kaloper\footnote{E-mail: 
kaloper@hepvs6.physics.mcgill.ca}\\
\vspace*{0.2cm}
{\it Department of Physics, University of Waterloo}\\
{\it Waterloo, ON N2L 3G1, Canada}\\
\vspace*{0.4cm}
Keith A. Olive\footnote{E-mail: olive@mnhep.hep.umn.edu}\\
\vspace*{0.2cm}
{\it School of Physics and Astronomy, 
University of Minnesota}\\
{\it Minneapolis, MN 55455, USA}\\
\vspace{2cm}
ABSTRACT
\end{center}
In this article, we examine the possibility
that there exist
special scalar-tensor theories of gravity 
with completely nonsingular
FRW solutions. Our investigation 
in fact shows that while 
most probes living in such a 
Universe never see the singularity, 
gravity waves always do. 
This is because they
couple to both the metric and 
the scalar field, in a way 
which effectively forces them 
to move along null geodesics of 
the Einstein conformal frame. 
Since the metric of the Einstein
conformal frame is always singular 
for configurations where
matter satisfies the energy 
conditions, the gravity wave world
lines are past inextendable 
beyond the Einstein frame
singularity, and hence the 
geometry is still incomplete,
and thus singular. We conclude that the 
singularity cannot be entirely removed, 
but only be made invisible 
to most, but not all, 
probes in the theory.

\vfill
\setcounter{page}{0}
\setcounter{footnote}{0}

\section{Introduction}

One of the longest-standing challenges to 
our understanding of gravity has been the 
singularity problem. In any generic theory of gravity 
and matter, under reasonable assumptions about 
the interactions between particles and fields 
and about the ways of communicating these
interactions (which translate into, at least
classically, fairly loose energy conditions), 
produces solutions which contain maelstormian 
regions of unbounded curvature \cite{he}. 
In this sense, such solutions 
seem to point to an intrinsic deficiency of the theory 
that gave them birth, because the very theory that 
predicted such maelstorms looses meaning in these
limits. Yet, we must note that not all is ill with
the fact that such strongly coupled regimes are generic 
in classical gravitational theories.
Because singular regions generally involve very strong
forces between particles and very high energies, 
close to a singularity much of the observed matter
structure in the Universe can be created starting
from arbitrary initial conditions.
Further, this mechanism is built naturally
into the theory, such that extrapolating present 
conditions backwards inevitably results in circumstances 
under which the present could be shaped in a generic way. 
Indeed, the vast body of astrophysical observation 
does indicate that such a dramatic event, the Big Bang, did 
take place in the past. The cosmological singularities
in the past, then, seem perfectly suited to encode such 
cosmogonic furnaces into the theory\footnote{Aside from 
the cosmological arguments 
favoring the presence of singularities in any theory
of gravity, which we describe in the text, there also
exist more theoretically-minded arguments, which suggest
that singularities may be needed in any theory of gravity 
for consistency reasons. Specifically, G. Horowitz
and R. Myers \cite{hm} have recently 
proposed that naked singularities 
in a theory of gravity are needed to separate positive mass
black hole solutions from the negative mass ones. 
If the negative mass solutions were not singular, 
they would be more energetically favorable than 
the empty space, thus yielding vacuum instability in
any quantum theory one might attempt to build
on the original classical theory.}. 
Our task, hence, seems to be to bring into accord 
the facts and the fiction by constructing a theory which 
could encompass all the features of the Big Bang while
hopefully not incorporating its own demise, in the form
of an uncontested singularity.

While it is 
not known how, and even if, all singularities may be 
regulated in a theory of gravity,
we may be able to gather interesting information
by studying the difficulties which emerge in the attempt
to smooth the edges of the Universe in the existing
models. A start for any such investigation, of 
course, should be Einstein's theory of General 
Relativity, since it is in an amazingly good 
agreement with experimental observations. In this
theory, however, the general theorems due to 
Hawking and Penrose show that the configurations 
which are determined by the coupled
Einstein-matter equations of motion, 
under the assumption that the above-mentioned 
energy conditions are valid, always contain
singular regions \cite{he}. The singularities in this 
context are signaled by geodesic incompleteness. 
This means that particle trajectories 
in such geometries cannot be continued 
past some hypersurface, because they get so 
strongly focused by gravity's pull that 
they begin to intersect and hence are not 
smooth any more. This phenomenon does not 
always imply that the curvature of 
the metric grows beyond bound in such limits. 
However, the geodesic curvatures do, since the 
geodesic bundles get squeezed very tightly. 
Hence an observer seeing such geodesic curvatures 
would indeed see a very strong force. 
With this in 
mind, the notion of geodesic incompleteness is indeed 
a good indicator of the singularity in the space-time.

The presence of singularities in General Relativity 
can be taken as a sign that at very high energies, 
or equivalently at very short distances, the theory 
fails to be a completely consistent description of 
Nature. 
Because at such high energies quantum effects
become significant, this suggests that General Relativity 
should be superseeded by the quantum 
theory of gravity, within which we should seek the 
answer to the conundrum of singularities. Whereas we 
are still lacking a complete formulation of the 
quantum theory of gravity, there at present is at 
least one very strong contender, string theory. 
Although the recent developments in string theory, 
assisted by the discovery of the 
power of duality \cite{witten}, 
have greatly improved our understanding of it, the 
theory is still not known in a way that would 
enable us to ask the questions about space-time in 
a general manner. Instead, we have to either resort 
to the effective action approach which takes into
account stringy phenomena in perturbation theory \cite{act}, 
or we could study some special classes of string 
solutions which can be formulated in the nonperturbative 
regime \cite{kk}. The latter approach is clearly more powerful 
in that it allows us to investigate more thoroughly the
quantum dynamics of the system under scrutiny. But 
this is available only for some special solutions, 
most notably the BPS states in the string spectrum, and not 
for any solution we might be interested in. In particular, 
there still does not exist a nonperturbative formulation 
of generic cosmological solutions in string theory. Hence 
all the investigations of ``realistic" string cosmologies 
have been carried out essentially in the effective action 
approach, which is valid for the weak to medium range of 
couplings and curvatures 
(see, for example, \cite{bd} - \cite{mology}).
We will not dwell on the details of these investigations here, 
other than to mention that the departure of string-theoretic 
solutions away from General Relativity is induced 
by the presence of additional degrees of freedom which arise 
in the massless string spectrum. These fields, the scalar 
dilaton field, the torsion tensor (or Kalb-Ramond) field, 
and others, couple to each other and to gravity nonminimally, 
and can influence the dynamics significantly.
Thus it makes sense to ask if the spectrum of the theory
may be tailored in such a way as to produce
solutions which do not feature any singular 
behavior \cite{gv} - \cite{kmo},
even if that means abandoning string theory and just 
constructing certain toy models for the purpose of studying 
the strong coupling limit. 

This approach has been taken recently in \cite{Bar,Rama}.
There a special class of scalar-tensor theories of gravity 
was considered, where the matter 
does not couple to the scalar field in some generalized
Jordan-Brans-Dicke (JBD) conformal frame, but the JBD 
coupling of the scalar to the metric depends on the value 
of the scalar field. It turns out that by choosing the JBD 
coupling function one could construct the metric in the JBD frame 
which is smooth over an infinite interval of the JBD comoving time. 
Since the matter fields couple only to the JBD frame metric, 
they don't feel any singularity at all. However, in the 
Einstein (E) conformal frame, where the metric degrees of 
freedom have a canonical kinetic term, if the
matter sources satisfy the energy conditions and 
consequently the dynamics are still subject to 
Hawking-Penrose theorems, the curvature 
singularities still plague all the solutions.
The E frame singularities are invisible to matter
probes which do not move along E frame geodesics. 
Rather, in the E frame matter couples to the scalar field, 
and this coupling is adjusted precisely to even out any bumps in the metric. 
If there were any type of probe which 
couples both to the JBD scalar and the JBD metric, 
it would detect the original singularity. 
The probe would reach the singularity 
in a finite extent of its world-line, 
after which it could not
be extended any further. This would signal that the geometry
is still incomplete, at least from the point of view
of certain observers.  Once the subleading 
interactions are properly taken into account, these observers
can communicate the presence of the singularity to the remainder. 
Hence, to check whether a solution of some nonminimal theory
of gravity is singular or not, it is not sufficient to show that
curvature invariants are finite in some chosen conformal frame
and that the space-time geometry is complete in it. 
Instead, we must investigate world-lines of all the physical probes
in the theory. Only if all these world-lines
are not inextendable can the space-time be complete and hence
nonsingular. As long as there is even a
single type of excitation which sees the singularity in any frame, 
the singularity is not absent, but lurks in the geometry waiting 
to exert its influence on the theory.

Our purpose here is to show that 
there is at least one such degree of freedom 
in all the scalar-tensor models studied in \cite{Bar,Rama}. 
It is the graviton itself. Being generic and model-independent,
and not immune to the singularity, it resuscitates the singularity
back into existence of any realistic observer in such theories. 
To demonstrate this, we only need to look at the classical
theory. We will present the equations of motion of the genuine
gravity waves, i.e. the tensor perturbations of the metric, and
take the geometrical optics limit to show that the wave packets of
gravity waves move along the E frame null geodesics.
These trajectories are past inextendable because of the singularity,
and the graviton wave packets reach the singular hypersurface
after a finite extent of the affine parameter along their
world-lines. Given this, the singularity cannot be completely removed
from the geometry: the gravity waves can communicate its presence to 
all other degrees of freedom in a finite time, ultimately 
making the singularity observable. 
The paper is organized as follows. In the next section, we will
review the JBD models studied in \cite{Bar,Rama}, and establish the
relationship between the JBD and E frames. Section 3. is 
devoted to the derivation of the gravity wave equations of
motion, both in the JBD and E frames. The singularity is the
central notion of section 4. There we will carry out the geometrical 
optics approximation and derive the gravity wave world-lines.
In the last section, we will give our conclusions.

\section{Hiding the Singularity}

Here we will review the models and 
solutions studied in \cite{Bar,Rama}
in order to set the stage for out 
investigation. The theories 
investigated in these articles were defined by the 
JBD actions with a variable 
parameter $\omega(\chi)$, given as
\be
\label{JBDact}
S=\int d^4x \sqrt{g} \Bigl\{\chi R 
- \frac{\omega(\chi)}{\chi} (\nabla \chi)^2
- {\cal L}_m({\cal Y}, \nabla {\cal Y}, g_{\mu\nu}) \Bigr\}
\ee
where $\chi$ is the scalar JBD field, 
$R$ is the scalar curvature of
the JBD frame metric $g_{\mu\nu}$ and 
${\cal Y}$ and $\nabla {\cal Y}$ are any
other matter degrees of freedom 
and their derivatives (field strengths).
Our signature conventions are
$g_{\mu\nu}=(-,+,+,+)$ and 
$R^{\mu}{}_{\nu\lambda\sigma} = \partial_\lambda
\Gamma^{\mu}_{\nu\sigma} - ...$.
Note that in order to ensure that 
the gravitational degrees of freedom 
are not ghost-like (i.e. that the 
graviton propagator never has negative residue),
we must require $\chi \ge 0$. 
For simplicity, we choose units such that in the
E frame we have $16\pi G_N=1$.
The approach taken in \cite{Bar,Rama} was to 
specify the function $\omega(\chi)$
in attempting to remove curvature 
singularities in spatially flat
FRW solutions. Note that in the JBD frame, the matter 
fields do not couple to the
scalar $\chi$ at the tree level. To get the equations 
of motion, we can simply vary the
action with respect to the independent 
degrees of freedom $g_{\mu\nu}$,
$\chi$ and ${\cal Y}$. After a bit of 
straightforward algebra, we find \cite{wet}
\ba
\label{JBDeom}
G_{\mu\nu}
&=& \frac{1}{\chi} \nabla_{\mu} \nabla_{\nu} \chi
- \frac{1}{\chi} g_{\mu\nu} \nabla^2 \chi
+ \frac{\omega}{\chi^2} \Bigl(\nabla_{\mu} 
\chi \nabla_{\nu} \chi
- \frac{1}{2} g_{\mu\nu} (\nabla \chi)^2 \Bigr) 
+ \frac{1}{2\chi} T_{\mu\nu} \nonumber \\
&&2 \nabla_{\mu} \Bigl( \frac{\omega}{\chi} 
\nabla^{\mu} \chi \Bigr) + R
- \frac{d}{d\chi}\Bigl(\frac{\omega}{\chi}\Bigr)
(\nabla \chi)^2 = 0 \\
&& \nabla_{\mu} T^{\mu\nu} = 0 ~~~~~~~~~~~~~~~
T_{\mu\nu} = -2\frac{\delta {\cal L}_m}
{\delta g^{\mu\nu}} - g_{\mu\nu}
{\cal L}_m \nonumber
\ea
where $G_{\mu\nu} = R_{\mu\nu} - \frac{1}{2} 
g_{\mu\nu} R$ is the Einstein tensor,
and $T_{\mu\nu}$ is the stress-energy tensor 
of the matter fields ${\cal Y}$. 
The conservation of the stress-energy 
$\nabla_{\mu} T^{\mu\nu}=0$ 
is equivalent to the matter
equations of motion $\nabla_{\mu} (\delta{\cal L}_m/ 
\delta \nabla_{\mu} {\cal Y}) =
\delta {\cal L}_m/ \delta {\cal Y}$. Let us now write 
the explicit form of (\ref{JBDeom}) for the spatially
flat FRW cosmologies, assuming that the matter
stress-energy tensor can be put in the perfect fluid
form, $T_{\mu\nu} = p g_{\mu\nu} + (p+\rho)u_\mu u_\nu$,
where $u^{\mu}$ is the velocity of the comoving
observer, $u_\mu u^\mu = -1$. The FRW metric is
\be
\label{FRWJBD}
ds^2 = - d\tau^2 + a^2(\tau) d\vec x^2
\ee
and the comoving velocity is $u^{\mu}={\rm diag}(1, \vec 0)$.
The equations of motion (\ref{JBDeom}) become \cite{wet}
\ba
\label{JBDefrw}
&&3 H^2 = \frac{\omega}{2}\frac{\chi'^2}{\chi^2} 
- 3 H \frac{\chi'}{\chi} + \frac{\rho}{2\chi} ~~~~~~~~~~~
\rho' + 3 H (\rho +p)=0 \nonumber \\
&&~~~~~~2H' + 3H^2 + \frac{\chi''}{\chi} + 2 H \frac{\chi'}{\chi} 
+\frac{\omega}{2}\frac{\chi'^2}{\chi^2} + \frac{p}{2\chi} = 0 \\
&&~~~~~~2\frac{\omega}{\chi}\Bigl(\chi'' + 3H \chi'\Bigr)
+\frac{d}{d\chi}\Bigl(\frac{\omega}{\chi}\Bigr) \chi'^2 
= 6H'+12H^2 \nonumber
\ea
The primes denote derivatives with respect to the JBD frame
comoving time $\tau$, and the JBD frame Hubble parameter is
$H=a'/a$. For simplicity's sake, we will assume that
the equation of state is $p=\gamma \rho$, with $\gamma$
a constant. This is not true in the real world, since 
we know that $\gamma$ must be a function of the temperature
of the Universe, and hence of time. However, since we
want to investigate the possibility of constructing 
a classical nonsingular Universe solution, this is a 
sufficiently good approximation to start with.
We will determine the limits
on $\gamma$ later.

Instead of deriving the equations of 
motion (\ref{JBDeom}) in the JBD frame,
we could have equally well 
used the E conformal frame. 
Conformal transformations of 
the degrees of freedom in a specified
theory are merely field redefinitions,
and so they cannot change physics \cite{redefs}.
In this sense, they are to be regarded 
as changes of reference frames,
which leave physical observables 
invariant. The important issue here
then is clearly to identify the 
observables correctly. To do that,
we must specify an observer, building 
it out of the physical fields in the theory.
Once given, this observer doesn't 
care which reference frame we use 
to compute observables with. So, if 
we define the E frame metric and
scalar field by
\be \label{conft}
\bar g_{\mu\nu} = \chi g_{\mu\nu} ~~~~~~ 
\phi = \int d\chi \frac{\sqrt{2\omega + 3}}
{\sqrt{2} \chi} 
\ee
(using the overbar to distinguish 
between the two conformal frames) 
and, following \cite{Rama}, assume that 
the function $\omega(\chi)$ is 
monotonic\footnote{We require that $\omega$ is
monotonic. But since 
$d\phi/d\chi = \sqrt{2\omega + 3}/\sqrt{2}\chi$,
and we wish that all curvature invariants, not only $R$ and
$R^{\mu\nu} R_{\mu\nu}$, are smooth, we should also require
that $\omega$ is analytic \cite{Rama}, 
because higher derivative
invariants depend on the derivatives of $\omega$. We will
discuss this in more detail following the equation
(\ref{RbarR}).}
such that the functional 
relationship $\phi = \phi(\chi)$ is 
everywhere invertible, we find 
that in the E frame the effective action is
\be
\label{Eact}
S=\int d^4x \sqrt{\bar g} \Bigl\{\bar R 
- (\bar \nabla \phi)^2
- \frac{1}{\chi^2(\phi)}{\cal L}_m({\cal Y}, 
\bar \nabla {\cal Y}, 
\frac{1}{\chi(\phi)} \bar g_{\mu\nu}) \Bigr\}
\ee
In this conformal frame, the 
metric kinetic term is canonical, i.e.
just $\bar R$, while now the 
matter fields ${\cal Y}$ couple {\it both}
to $\bar g_{\mu\nu}$ {\it and} 
$\phi$. Note that in order for (\ref{conft}) to be
well-defined, we must require $2\omega+3 >0$.
This is because when $2\omega+3=0$, the scalar
field is not dynamical, since a conformal 
transformation to the E frame removes its
kinetic term. Further, for $2\omega +3<0$,
the scalar would be ghost-like, since its kinetic
term would be negative.
Now, to find the equations of
motion in this frame, we can 
either take the equations (\ref{JBDeom})
and transform them to the E frame 
using (\ref{conft}), or we can vary the
action (\ref{Eact}). Since in the 
E frame the matter fields couple both
to the metric and the scalar $\phi$, 
the matter equations of motion are
a little bit more complicated than 
in (\ref{JBDeom}). Varying (\ref{Eact}) 
with respect to ${\cal Y}$, we obtain 
$\bar \nabla_{\mu} [(1/\chi^2) 
\delta{\cal L}_m/ \delta \bar \nabla_{\mu} 
{\cal Y}] = (1/\chi^2) \delta {\cal L}_m/ 
\delta {\cal Y}$. 
Furthermore, recalling that $\delta [(1/\chi^2) 
{\cal L}_m]/\delta \phi = -\{(d \chi/d\phi) 
[2{\cal L}_m + \bar g_{\mu\nu} 
(\delta {\cal L}_m/\delta \bar g_{\mu\nu})]\}/\chi^3 
= (d\chi/d\phi) \bar T/2\chi$, where $\bar T$ is the
trace of $\bar T_{\mu\nu}$, the resulting 
equations of motion in the E frame are \cite{wet,nko1}
\ba
\label{Eeom}
\bar G_{\mu\nu}
&=& \bar \nabla_{\mu} \phi 
\bar \nabla_{\nu} \phi
- \frac{1}{2} \bar g_{\mu\nu} 
(\bar \nabla \phi)^2 + \frac{1}{2}\bar T_{\mu\nu} 
~~~~~~~~~
 \bar \nabla^2 \phi = \frac{1}{4\chi} 
\frac{d \chi}{d\phi} \bar T \nonumber \\
\nabla_{\mu} T^{\mu\nu} &=& - 
\frac{1}{2\chi} \frac{d \chi}{d\phi} ~\bar T 
\bar \nabla^{\nu} \phi ~~~~~~~~~~~~~~~~~~~~
\bar T_{\mu\nu} = - \frac{2}{\chi^2} 
\frac{\delta {\cal L}_m}{\delta \bar g^{\mu\nu}} 
- \frac{1}{\chi^2} \bar g_{\mu\nu} {\cal L}_m 
\ea
Again, we will need (\ref{Eeom}) restricted on the
spatially flat FRW geometries with perfect fluid
sources, $\bar T_{\mu\nu} = \bar p \bar g_{\mu\nu}
+ (\bar \rho +\bar p) \bar u_{\mu} \bar u_{\nu}$,
where $\bar u^{\mu}$ is now the velocity of the 
E frame comoving observer.
The E frame metric is
\be
\label{efrwmet}
d\bar s^2 = - dt^2 + \bar a^2(t) d\vec x^2
\ee
and so $\bar u^{\mu} \bar u_{\mu} = -1$, $\bar u^{\mu} =
{\rm diag}(1,\vec 0)$. We will show that this is consistent with
the JBD frame comoving velocity shortly, by establishing
the transformation properties relating the JBD and E
quantities. The equations of
motion (\ref{Eeom}) then become \cite{wet,nko1}
\ba
\label{Efrwe}
3 \bar H^2 &=& \frac{\dot \phi^2}{2} + \frac{\bar \rho}{2} ~~~~~~~~
\dot {\bar \rho} + 3 \bar H (\bar \rho + \bar p) = 
\frac{\dot \phi}{2\chi} \frac{d\chi}{d\phi} 
(3\bar p - \bar \rho) \nonumber \\
&& ~~~~~~~~~~ \dot {\bar H} + \frac{\dot \phi^2}{2} 
+ \frac{\bar \rho + \bar p}{4} =0 \\
&& ~~~ \ddot \phi + 3 \bar H \dot \phi + \frac{1}{4\chi}
\frac{d\chi}{d\phi} (3\bar p - \bar \rho) = 0 \nonumber
\ea
Here the dot denotes derivatives with respect to
the E frame time $t$, and the E frame Hubble parameter
is $\bar H = \dot {\bar a}/\bar a$. 
Using the variational definition of the matter 
stress-energy tensor, we can 
immediately see that the conformal transformation
(\ref{conft}) induces the change of the
stress-energy tensor according to \cite{wet} 
$\bar T^{\mu}{}_{\nu} = 
T^{\mu}{}_{\nu}/\chi^2$. With this and (\ref{conft}), we 
can also easily show that the JBD frame equations
of motion (\ref{JBDeom}) and the E frame equations 
(\ref{Eeom}) map into each other. For the
variables describing the FRW universes (\ref{FRWJBD})
and (\ref{efrwmet}), the transformations are 
$d\tau = dt/\sqrt{\chi}$, $a(\tau) = \bar a(t)/\sqrt{\chi}$,
$p = \chi^2 \bar p$ and $\rho = \chi^2 \bar \rho$. 
To find out how the comoving velocity transforms, 
we should look at the comoving velocity vector fields 
in each frame. In the JBD frame, we have 
$U = \partial_{\tau}$ and in the E frame 
$\bar U = \partial_t$, so $U = \sqrt{\chi} \bar U$. 
But this means that the components of the comoving 
velocities in the two frames are identical, 
$u^{\mu} = \bar u^{\mu} = {\rm diag}(1,\vec 0)$,
as has been claimed above. This completes our 
survey of the conformal transformation rules for 
the quantities of interest here.

Now we can investigate the properties of
the theory, using either of the sets of
field equations (\ref{JBDeom})-(\ref{Eeom}). 
We should first recall 
the Hawking-Penrose singularity theorems \cite{he}.
To do this, we will use the E frame 
equations of motion (\ref{Eeom}), since
the gravitational equations of 
motion are the same as in General
Relativity. Consider any  
timelike geodesic with a
unit tangent vector field $\xi^{\mu}$
($\bar g_{\mu\nu}\xi^{\mu} \xi^{\nu} =-1$)
in a globally-hyperbolic 
space-time (i.e. space-time without any 
acausal pathologies such as closed time-like
curves, which we will assume here). The singularity
theorems then posit that these
geodesics are past-inextendable
(i.e. incomplete)
as long as the projection of the Ricci tensor on the
tangent vector field is positive semi-definite, 
$\bar R_{\mu\nu} \xi^{\mu} \xi^{\nu} \ge 0$. 
By the equations of motion, this condition 
can be recast as $\bar \Theta_{\mu\nu} 
\xi^{\mu}\xi^{\nu} + \bar \Theta \ge 0$, 
where the tensor $\bar \Theta_{\mu\nu}$ is the
total stress-energy, $\bar \Theta_{\mu\nu} =
\bar \nabla_{\mu} \phi \bar \nabla_{\nu} \phi
- (1/2) \bar g_{\mu\nu} (\bar \nabla \phi)^2 
+  \bar T_{\mu\nu}$. This requirement is 
called the strong energy condition (SEC), and 
is thought to be satisfied by most reasonable
classical matter sources. In terms of the 
principal values of the stress-energy
tensor $\bar \Theta^{\mu}{}_{\nu} = 
{\rm diag}(\hat \rho, \hat p_1, \hat p_2, \hat p_3)$, 
SEC translates into $\hat \rho + 
\Sigma^{3}_{i=1} \hat p_i \ge 0$, $\hat \rho + 
\hat p_i \ge 0, ~~i=1,2,3$. (Here we use the hat 
to distinguish the principal values of $\bar \Theta_{\mu\nu}$
from those of $\bar T_{\mu\nu}$, which are both defined
in the E frame.) The only feature of the JBD system we
will be investigating here is the effect of  
the variation of $\omega$ with 
$\chi$ on the singularities, which still needs to be 
specified. Therefore, we 
can assume that the matter fields ${\cal Y}$ 
obey SEC, meaning that $\bar T_{\mu\nu} \xi^{\mu} 
\xi^{\nu} + \bar T \ge 0$ for all timelike
geodesics $\xi$. For the homogeneous and isotropic cases
which we are interested in, this tells us that
$3\gamma +1 \ge 0$, or $\gamma \ge -1/3$.
When $\phi$ is included, its stress-energy
trivially satisfies SEC in the E frame. Hence, $\bar \Theta$ 
also does, being a linear combination of these 
two contributions. We see that 
in the E frame all timelike geodesics must be
past inextendable, and thus all cosmological
solutions are singular. 

Unfortunately, this 
does not specify the character of the singularity.
Generically, geodesic incompleteness
as an indicator of the presence of a singularity 
signals that the metric becomes
degenerate as some region of space-time 
is approached. 
To learn more about the actual properties of the
singularity, we must look at concrete 
solutions. 
If we limit our 
attention to spatially flat FRW cosmologies, 
which have matter sources that satisfy SEC,
we will typically find that the singularity
arises because at some time, say $t=0$, the
scale factor of the Universe vanishes (or diverges,
such as in pole-law inflationary solutions) 
as some power of $t$, $\bar a(t) \propto t^\alpha$,
$\alpha \ne 0$\footnote{It is clear that hypersurfaces
$t=const$ where $\bar a(t)$ is smooth and nonvanishing
cannot be singular, since the metric there is nondegenerate.
We are assuming that singularities 
can arise only as zeros or isolated 
singularities of $\bar a$ in 
the functional sense, and further 
that if $\bar a$ is unbounded
for some values of $t$, that such singularities are not
essential singularities - i.e. that $\bar a$ is analytic
everywhere near the singularities, and that if it diverges
there it admits a Laurent series expansion with a finite
number of divergent terms. This is consistent, because
essential singularities of $\bar a$ are never a part of the
manifold in the sense discussed in the text. For assume
$t=0$ were an essential singularity. Then the geodesic
distance from the singularity to anywhere else in the manifold
satisfies $\lim_{t\rightarrow 0} \Delta \bar \lambda 
\propto \lim_{t\rightarrow 0} t \bar a(t)$.
But since $t=0$ is an essential singularity, $\lim_{t\rightarrow 0}
t \bar a \rightarrow \infty$. Hence, $\Delta \bar \lambda$ always
diverges.}.
Since the curvature scalar can
be expressed as $\bar R = 6 \dot {\bar H} + 12 \bar H^2$,
we find that in the limit $t \rightarrow 0$, 
$\bar R = 6\alpha(2\alpha-1)/t^2
+ subleading ~terms$\footnote{Note that this is valid
even for the cases when the scale factor vanishes faster
than the power law, such as 
$\bar a \propto t^{2/3} (\ln(t))^{1/2}$,
which is known to be the limiting form of the solution
dominated by composite hadrons \cite{comp}.}. 
So for $\alpha \ne 1/2$, 
the scalar curvature diverges at $t=0$. If 
$\alpha =1/2$, then near the singularity 
the Universe is radiation-dominated. The
scalar curvature vanishes because the 
radiation sources are conformally invariant
and hence have vanishing trace of the stress-energy
tensor. However, the square of the Ricci tensor
then diverges, as $\bar R_{\mu\nu} 
\bar R^{\mu\nu} \propto 1/t^4$. Moreover, if
we look at causal geodesics, we can see that
in order to move between times $t$ and $t=0$,
they require a lapse of the affine
parameter equal to 
\be
\label{affineE}
\Delta \bar \lambda =
\int^t_0 dt ~\bar a/\sqrt{v^2 + m^2 \bar a^2}
\ee
which can be obtained by solving the causal geodesic
equations for the metric (\ref{efrwmet}), and where $v^2$
and $m^2$ arise as constants of integration;
$v^2$ is a nonnegative constant, and $m^2 = 0,1$. 
In the limit of small $t$,
when $\bar a 
\propto t^\alpha$, we can approximate this 
expression with $\Delta \bar \lambda \propto 
t^{1+\alpha}/\sqrt{v^2 + m^2 t^{2\alpha}}$, hence
noting that we can always choose $v^2$ and $m^2$ 
such that $\Delta \bar \lambda$ is finite\footnote{e.g. for
$\alpha >-1$, take $m^2=0$ and $v^2=1$; then, 
$\Delta \bar \lambda \propto t^{1+\alpha}$; 
if $\alpha \le -1$, choose $v^2 =0$, 
$m^2=1$ (static observers!),
so that $\Delta \bar \lambda \propto t$.}. This
means that there always exist causal
geodesics which reach out of the singularity
to anywhere in the Universe in a finite
proper time - i.e. they are past-inextendable
and so incomplete. As a result, all the FRW 
solutions with this type of behavior are 
indeed singular.

To see how the mechanism of conformal
transformations could regulate the singularities,
recall that since all FRW solutions (that is,
$k=\pm 1,0$) are conformal 
to static geometries with maximally 
symmetric spatial slices, the Weyl tensor of 
these solutions is a constant and hence does not 
encode any information about the
Big Bang singularity. The information about 
the singularity is completely encoded in 
the Ricci sector of the curvature, which 
changes under conformal transformations.
Indeed, we can look at the leading form
of the curvature in the vicinity of 
the singularity of every solution 
we have discussed above. Applying the
field redefinitions (\ref{conft}) to the
solutions, and using the equations of
motion in the E frame (\ref{Eeom}), we 
can show that the JBD and E frame curvatures 
are related by
\be
\label{RbarR}
R=\chi\Bigl\{\bar R + \frac{3\bar T}{4\omega +6} 
- \frac{3 (\bar \nabla \phi)^2}{(2\omega +3)^2} 
\Bigl(2\chi \frac{d\omega}{d\chi} + 2\omega +3\Bigr)\Bigr\}
\ee
We have seen above that as the singularity is
approached, the E frame Ricci scalar diverges
as $\bar R \propto 1/t^2$. Hence if in this limit
$\chi \rightarrow t^\beta$ with $\beta \ge 2$,
the divergent contribution of $\bar R$ in 
(\ref{RbarR}) can be tamed. What about the other two 
terms? It is straightforward to see that in the
limit $\bar a(t) \rightarrow t^\alpha$, the 
equations of motion (\ref{Efrwe}) are approximated by 
\ba
\label{nseom}
3\frac{\alpha^2}{t^2} = \frac{\dot \phi^2}{2} 
+ \frac{\bar \rho}{2} && 
~~~~~~~~~~~~ \frac{\alpha}{t^2} = \frac{\dot \phi^2}{2} +
\frac{1+\gamma}{4} \bar \rho \nonumber \\ 
\dot{\bar \rho} + 3(1+ \gamma)\frac{\alpha}{t} 
\bar \rho - A \dot \phi \bar \rho &=& 0 ~~~~~~~~~~~ 
\ddot \phi + 3\frac{\alpha}{t} \dot \phi 
+ \frac{A}{2} \bar \rho = 0 
\ea
where $A = (3\gamma -1)/\sqrt{4\omega(\chi_0) +6} \ge 0$ when
$\gamma \ge -1/3$ is a finite constant since $2\omega+3>0$. 
On the other hand, we can see from (\ref{JBDeom}) that
we can always write down the exact solution for the
fluid, since in the JBD frame the fluid couples only to
the metric. The solution is $\rho = \rho_0/a^{3(1+\gamma)}$.
Now, since $\bar \rho = \rho/\chi^2$, we can transform 
the energy density to the E frame: $\bar \rho = \rho_0 
\chi^{(3\gamma -1)/2}/\bar a^{3(1+\gamma)}$. In
the limit $t \rightarrow 0$, as $\bar a \propto t^\alpha$
and $\chi \propto t^\beta$, we find 
$\bar \rho \propto t^{(3\gamma -1)\beta/2 
- 3\alpha(1+\gamma)}$.
Hence in order for the $\rho$ pole contribution
to the JBD curvature to be smoothed, we must require
$\beta + (3\gamma -1)\beta/2 - 3\alpha(1+\gamma)\ge 0$.
From the equations of motion near the
singularity, we see that $\bar \rho \le 6 \alpha^2/t^2$
for all $t$. Therefore, for all solutions,
$(3\gamma -1)\beta/2 - 3\alpha(1+\gamma) \ge -2$,
which means that the regulator field $\chi$ is no worse than in 
the previous case. Let us now compare this to the 
scalar field contributions. Assuming a strictly greater than
order relation in the previous inequality, so that $\bar \rho$
falls off slower than $1/t^2$, 
we see from the constraint equation
solved for $\dot \phi$, $\dot \phi^2/2=3\bar H^2 - \bar \rho$,
that near the singularity
the dominant contribution comes from $\bar H$. This means that
the scalar field is $\dot \phi \propto \sqrt{6}\alpha/t$. 
But then, sufficiently
close to $t=0$, the $\bar \rho$-dependent contribution 
to the $\ddot \phi$ equation is also negligible. As
a result, in this limit $\alpha \rightarrow 1/3$,
and hence the solution approaches
$\bar a \rightarrow t^{1/3}$, $\phi \rightarrow \phi_0 +
\sqrt{2/3} \ln(t)$ with negligible matter field 
contributions - i.e. we get the scalar field-dominated cosmology 
\cite{nko1,tsey}.
Inspecting (\ref{nseom}) near the singularity we can see 
that the matter sources can never 
dominate over the scalar 
field\footnote{Suppose they did; then, near
the singularity the scale factor would behave as 
$\bar a \propto t^{2/3(1+\gamma)}$ and hence 
$\bar \rho \propto 1/t^2$.
But this then produces the response in the scalar field
according to $\dot \phi = P/t^{2/1+\gamma} + Q/t$, where 
$P$ and $Q\ne0$ are integration constants.}. Hence near the
singularity, the $\phi$-dependent contributions can 
never be subleading to the matter stress-energy contribution, 
regardless of the kind of matter field. 

The only remaining possibility
is that the two sources remain of equal importance near the
singularity. Indeed, a glance at (\ref{nseom}) suggests that
a solution of the asymptotic form $\dot \phi \propto 1/t$,
$\bar \rho \propto 1/t^2$ near the singularity is admissible.
Yet, a closer look reveals that, unless $\gamma=-1$,
the four equations in (\ref{nseom}) are consistent only if 
$\bar \rho=0$ - hence again leading to the scalar 
field-dominated solution. If $\gamma=-1$ (which corresponds
to the JBD frame cosmological constant, since $\rho = const$
from the equations of motion
(\ref{JBDefrw})), the solution which treats the sources 
in an egalitarian way is admissible. However, in this case
the parameter $\beta$ cannot be adjusted to be $\ge 2$. 
It is fixed by $\gamma =-1$ and $3\alpha(1+\gamma) -
\beta(3\gamma-1)/2 =2$ to be $\beta=1$. So the
JBD frame solution is still singular! For this case, it is
impossible to remove the E frame singularity by going to
the JBD frame via the conformal transformation (\ref{conft}),
regardless of the form of the coupling 
function $\omega(\chi)$\footnote{For completeness, we 
should also mention another special example where frame 
switching fails, although this case does not violate the 
results of \cite{Bar,Rama}, since the scalar field is 
constant. Namely, if the matter is in the form of pure radiation,
there exist solutions with $\phi=const$, as can be seen
from either (\ref{JBDeom}) or (\ref{Eeom}). Hence
the JBD and the E frame are identical, and so the  
singularity is not removed. Nevertheless, 
these solutions are not past 
attractors for all late-time radiation-filled Universes,
since as long as $\dot \phi \ne 0$, early enough it will
dominate over radiation. So these solutions are not generic,
but they illustrate that scalar field must dominate
near the singularity in order for the frame switching to
be successful in smoothing the solution.}.
From this discussion, we can see that all the JBD 
frame solutions without a 
singularity must have a unique behavior near the singularity 
in the E frame - that of the scalar field-dominated cosmology. 
In fact, this is the reason why a conformal transformation
can be used to smooth the solution. Once the solution is
dominated by the scalar field near the singularity, we
can choose an almost arbitrary function of the scalar 
field to conformally transform the solution with - as long
as the conformal factor vanishes in the limit $t\rightarrow 0$
at least as fast as $t^2$.

To complete the smoothing of the singularity within
the context of the JBD theory, we must require that the
hypersurface $t=0$ can never be reached by any particle
in the JBD world. 
If we look at the matter sector of the theory,
in which fields move along the JBD frame geodesics, we should
require that no such geodesic ever reaches the 
$t=0$ hypersurface. 
When we 
solve the geodesic equations of the JBD metric 
(\ref{FRWJBD}), and express them in terms of the
E frame time for convenience, we find that the
JBD frame geodesic lapse between hypersurfaces 
$t_1$ and $t_0$ is 
\be
\label{JBDaffine}
\Delta \lambda = \int^{t_1}_{t_0}
\bar a dt/\sqrt{v^2 \chi^2 + m^2 \bar a^2 \chi}
\ee
where as before $v^2\ge 0$ and $m^2=0,1$ are integration 
constants. To make sure that all the geodesics are complete,
we must require that $\Delta \lambda$ diverges as either
of the limits of integration goes to zero. Substituting
the asymptotic form of the functions $\bar a$ and $\chi$
which we have deduced above, we find that as long as
$\beta \ge {\rm max}(2, \alpha+1)$, the geodesic 
distance $\Delta \lambda$ between the $t=0$ hypersurface
and any other $t=const$ hypersurface diverges. Further,
if we set $v^2=0$, $m^2=1$, we recover the integrated
coordinate transformation between the JBD comoving
time $\tau$ and the E comoving time $t$, 
$\tau = \int^{t} dt/\chi$, as we should. So we see
that the hypersurface $t=0$ as seen from the 
JBD frame corresponds to the infinite past 
(or future!) of the solution, as long as $\beta \ge 2$:
$\lim_{t\rightarrow 0} \tau \rightarrow \pm \infty$.
Therefore, the conformal transformation to the JBD frame 
does push away the singularity to an infinite distance,
while making the JBD curvature finite. 

At this point, we must remember that while we are
using the conformal transformation (\ref{conft})
to regulate the solution at $t=0$, we must make sure
that it does not behave badly elsewhere. 
The conditions which $\omega(\phi)$ must satisfy in order
to guarantee that the JBD solutions are nonsingular can be
summarized in terms of $\Omega =2\omega +3$, 
following \cite{Rama},
as 

\noindent (1) $d^n \Omega/d\chi^n$ are smooth 
everywhere except in the
limit $\chi \rightarrow 1$ for all $n \in {\bf Z_+}$, 

\noindent (2) as $\chi \rightarrow 1$, both 
$\Omega$ and it's derivatives diverge as some inverse power of
$(1-\chi)$, and 

\noindent (3) finally and most importantly, that near the
singularity the solutions are dominated by the JBD scalar 
field $\phi$ (or equivalently, $\chi$), which implies
$\lim_{\chi\rightarrow 0} \Omega \ge 1/3$. 

\noindent We emphasize here
that whereas this condition is vacuous for a range of admissible
types of matter, it is not automatically true for all solutions. 
In particular, we have given two simple examples (with
the JBD frame matter being either the cosmological constant
or a radiation fluid) where the scalar field in the E frame
is not dominant near the singularity and hence the JBD frame
solutions remain singular. 

However, as we have indicated earlier 
in the discussion preceding
the investigation of the 
conformal pictures of the physics
given by (\ref{JBDact})-(\ref{Eact}), 
even if the conformal rescaling 
did produce a smooth metric in the 
JBD frame, it may not have really
removed the singularity if there 
still remained even a single 
degree of freedom which coupled 
to the E frame geometry. 
In what follows we are
going to show that
gravitational waves 
move along worldlines in the 
JBD frame which are not geodesics.
Rather, they are deformed by 
the dilaton force such that they 
are identical to the E frame geodesics! 
This should not be a complete 
surprise since it is the E frame where 
the graviton kinetic term takes
the canonical form, being just 
the Ricci scalar. It has been argued
by Shapere, Trivedi, and Wilczek \cite{STW}
that in theories of antisymmetric $p$-forms
coupled to scalars via 
$f(\phi) F^2_{\mu_1 ... \mu_{p+1}}$ the
tensor field quanta move along 
geodesics of the metric where the
term $f(\phi)$ is absorbed away 
by a conformal transformation. 
This was later proved in \cite{ckmo} 
by the present authors and
collaborators to hold as the 
geometrical optics limit of the 
$p$-form field equations. Since 
gravity is a gauge theory like
the $p$-form field theories, 
when we apply the same technique to
the equations of motion of gravitational 
perturbations in the FRW
background, we find that the gravity 
waves in general see a scalar force. 
Therefore since the E frame metric is 
singular, the gravity wave 
worldlines are past-inextendable,
and hence the full physical 
arena of the theory given in either
frame is still incomplete! 
This must be taken as a signature of a latent
singularity, which simply didn't 
go away by a conformal transformation,
but only appeared invisible 
to the matter sector probes. 

\section{Gravity's Redoubt}

Below we will consider small 
perturbations away from a fixed curved 
background, and look at their 
dynamics to the lowest order in the 
fluctuation. Since we want to study only
the pure gravitational excitations, we will assume that the
matter and the JBD scalar are unperturbed, and impose the 
transverse traceless gauge on the metric fluctuation.
Further, since we will look for the perturbations around
the FRW solutions, we will impose the stationary gauge on the
background metric, which will therefore leave the background
solutions identical to the ones 
studied in section 2. The resulting 
equations of motion give the correct 
dynamics of the two independent
graviton polarizations, the $+$ 
and the $\times$ modes, in exactly
the same way as discussed previously 
\cite{ka,gas}. In the E frame,
these equations will be the same as 
in the ordinary General Relativity,
because of the gauge conditions. 
In the JBD frame, they will contain
an additional coupling to the 
scalar $\chi$, which arises from the
noncanonical form of the graviton propagator.

In the E frame, therefore, we will use 
the following ansatz for the metric
and matter fields:
\be
\label{Epert}
\hat {\bar g}_{\mu\nu} = \bar g_{\mu\nu} + \bar h_{\mu\nu}
~~~~~~~~~ \delta \phi = \delta {\cal Y} = 0
\ee
The genuine gravitational 
degrees of freedom correspond to two 
graviton polarizations defined by imposing the 
transverse traceless conditions 
on the metric perturbation.
In the momentum space, 
this means that we require
the polarization tensor of the 
graviton $\bar \epsilon_{\mu\nu}$
to be traceless and orthogonal 
to the direction of motion,
i.e. $\bar \epsilon^{\mu}{}_{\mu} 
= k^{\mu}\bar \epsilon_{\mu\nu}=0$.
In terms of the perturbation 
$\bar h_{\mu\nu}$, these conditions
can be written down as 
$\bar h^{\mu}{}_{\mu} = \bar \nabla_{\mu}
\bar h^{\mu}{}_{\nu} = 0$. Here 
the covariant derivatives and raising and 
lowering of indices are taken 
with respect to the background metrics
$\bar g_{\mu\nu}$. The inverse 
metric, to linear order in
perturbation $\bar h$ is 
$\hat{\bar g}^{\mu\nu} = \bar g^{\mu\nu} - 
\bar h^{\mu\nu}$. Notice here 
that the determinant of the metric
is not perturbed: $\hat{\bar g} 
=\det(\hat{\bar g}_{\mu\nu})
= \bar g + \bar h^{\mu}{}_{\mu} = \bar g$.
A straightforward calculation shows that
the Christoffel symbols in 
the perturbed background are
$\hat{\bar \Gamma}^{\mu}_{\nu\lambda}
 = \bar \Gamma^{\mu}_{\nu\lambda}
+ \bar \gamma^{\mu}_{\nu\lambda}$ where 
$\bar \gamma^{\mu}_{\nu\lambda} = (1/2) 
\bar g^{\mu\rho}
(\bar \nabla_{\lambda} \bar h_{\nu\rho} 
+ \bar \nabla_{\nu} \bar h_{\rho\lambda}
- \bar \nabla_{\rho} \bar h_{\nu\lambda})$.
The Ricci tensor, expanded to linear order in
$\bar h_{\mu\nu}$, is, in 
the transverse traceless gauge,
\be
\label{Erich}
\bar {\cal R}_{\mu\nu} = \bar R_{\mu\nu} + 
\frac{1}{2} \bar \nabla_{\lambda} \bar \nabla_{\mu} 
\bar h^{\lambda}{}_{\nu} +
\frac{1}{2} \bar \nabla_{\lambda} \bar \nabla_{\nu} 
\bar h^{\lambda}{}_{\mu} - 
\frac{1}{2} \bar \nabla^2 \bar h_{\mu\nu}
\ee
Note that the perturbation in 
$R_{\mu\nu}$ is traceless because of the
radiation gauge conditions, as expected.
Tracing out the gravitational 
equations of motion in
(\ref{Eeom}) and subtracting 
the Ricci scalar-dependent term,
we see that 
$\bar R_{\mu\nu} = \bar \nabla_{\mu} \phi
\bar \nabla_{\nu} \phi + \bar T_{\mu\nu} - 
(1/2) \bar g_{\mu\nu} \bar T$. The perturbed equations
of motion are identical to these. In order to find the
equations of motion of the perturbations, then, we have to 
expand the equations to linear order in $\bar h_{\mu\nu}$
and cancel the lowest order terms since the background
is a solution to (\ref{Eeom}). We also need to demonstrate
that the perturbation is consistent with the matter sector
of the theory, in that it is not exciting any perturbations
there. At this time, it is convenient to refer to the
explicit form of the background solution, and introduce
the gauge conditions for it. We will work with the
synchronous gauge, where the metric is exactly the same
as in (\ref{efrwmet}), 
$d\bar s^2 = -dt^2 + \bar a^2(t) d\vec x^2$,
and the perturbation satisfies 
$\bar h_{00} = \bar h_{0k}=0$.
Then, on all cosmological 
backgrounds of the previous section,
$\bar h^{\mu\nu} \bar \nabla_{\nu} \phi 
= \dot \phi \bar h^{\mu0} =0$,
and so, the perturbed d'Alembertian 
is identical to the 
unperturbed one:
\be
\label{dal}
\hat {\bar \nabla}^2 \phi = \frac{1}{\sqrt{\hat{\bar g}}}
\partial_{\mu} (\sqrt{\hat{\bar g}} \hat{\bar g}^{\mu\nu}
\partial_{\nu} \phi) =
\bar \nabla^2 \phi - \frac{1}{\sqrt{{\bar g}}}
\partial_{\mu} (\sqrt{{\bar g}}{\bar h}^{\mu\nu}
\partial_{\nu} \phi) = \bar \nabla^2 \phi
\ee
Next, since we are working in the synchronous gauge, the
comoving velocity is unchanged: 
$\hat {\bar u}^{\mu} = \bar u^{\mu}=
{\rm diag}(1,\vec 0)$, and so is the trace of the
stress energy tensor. The perturbation of the 
stress-energy tensor for fluid sources is 
\ba
\label{pertt}
\delta {\cal T}_{\mu\nu} &=& 
\bar {\cal T}_{\mu\nu} - \frac{1}{2} 
\hat{\bar g}_{\mu\nu} \bar{\cal T}
- \bar T_{\mu\nu} + \frac{1}{2}
{\bar g}_{\mu\nu} \bar T \nonumber \\
&=& \bar p \bar h_{\mu\nu} - 
\frac{1}{2} \bar h_{\mu\nu} \bar T \\
&=& \bar h_{\mu\nu} 
\frac{\bar \rho + \bar p}{2} \nonumber
\ea
Also, the
stress-energy conservation equation is identical to the
case without the perturbation because of the
gauge conditions; specifically, $\hat{\bar \nabla}_{\mu}
\bar T^{\mu\nu} = {\bar \nabla}_{\mu} \bar T^{\mu\nu}$.
Substituting these conditions and the equation (\ref{Erich})
in (\ref{Eeom}), we obtain the 
covariant form of the equations 
of motion for tensor perturbations:
\be
\label{Epeom}
\bar \nabla_{\lambda} \bar \nabla_{\mu} 
\bar h^{\lambda}{}_{\nu} +
\bar \nabla_{\lambda} \bar \nabla_{\nu} 
\bar h^{\lambda}{}_{\mu} = \bar \nabla^2 \bar h_{\mu\nu}
+ \frac{\bar \rho + \bar p}{2} \bar h_{\mu\nu}
\ee
We now want to rewrite this equation in terms of the mixed
perturbation tensor $\bar h^{\mu}{}_{\nu}$, which is the
natural variable to use because of the gauge
conditions. Using the
explicit form of the background metric 
(\ref{efrwmet}), we can verify that in the synchronous
gauge the following conditions hold identically
(see, e.g. \cite{ka,gas}): 
\ba
\label{conds1}
&&\bar \nabla_{\nu} \bar h_{00} 
= \bar \nabla_{0} \bar h_{0\nu} = 0   ~~~~~~~~~~
\bar \nabla_j \bar h_{k0} = - \bar H \bar h_{jk} \nonumber \\
&&\bar \nabla_0 \bar h_{jk}= \dot {\bar h}_{jk} - 
2 \bar H \bar h_{jk} ~~~~~~~~~~ \bar \nabla_i \bar h_{jk} =
\partial_i \bar h_{jk}
\ea
Since these equations are covariant with respect to the
assumed background, we can raise 
the indices using the background
metric, and after some straightforward algebra, 
we can also verify the following conditions on the
second covariant derivatives of the perturbation \cite{ka,gas}:
\ba
\label{conds2}
&&\bar \nabla^2 \bar h^{0}{}_{0} = \bar \nabla^2 \bar h^{0}{}_{j} 
= \bar \nabla_{\mu} \bar \nabla^{0} \bar h^{\mu}{}_{0} 
= \bar \nabla_{\mu} \bar \nabla^{0} \bar h^{\mu}{}_{j} 
= \bar \nabla_{\mu} \bar \nabla^{j} 
\bar h^{\mu}{}_{0} = 0 \nonumber \\
&&~~~~~~~~~~~~~ \bar \nabla_{\mu} 
\bar \nabla^{j} \bar h^{\mu}{}_{k} =
(\dot{\bar H}+ 4\bar H^2) \bar h^{j}{}_{k} \\
&&\bar \nabla^2 \bar h^{j}{}_{k} 
= \frac{1}{\bar a^2} \vec \nabla ^2 
\bar h^{j}{}_{k} - \ddot {\bar h}^{j}{}_{k} - 
3 \bar H \dot {\bar h}^{j}{}_{k} 
+ 2 \bar H^2 \bar h^{j}{}_{k}
\nonumber
\ea
In these equations, $\vec \nabla^2$ 
is just the three-dimensional
flat space Laplacian, 
$\vec \nabla^2 = \Sigma^{3}_{j=1} \partial^2_j$.
Upon substituting these expressions in (\ref{Epeom}),
we find 
\be
\label{Epeex}
\ddot {\bar h}^{j}{}_{k} + 3 \bar H \dot {\bar h}^{j}{}_{k}
- \frac{1}{\bar a^2} \vec \nabla^2 \bar h^{j}{}_{k} 
+ (2\dot {\bar H} + 6 \bar H^2 + \frac{\bar p - \bar \rho}{2}) 
\bar h^{j}{}_{k} = 0
\ee
The last term looks like the environment-induced mass; however, 
by the E frame FRW equations of motion (\ref{Efrwe}), this term is 
identically zero. Hence, finally, the equations of motion for the
transverse traceless perturbations of the metric are
\be
\label{Epeexfin}
\ddot {\bar h}^{j}{}_{k} + 3 \bar H \dot {\bar h}^{j}{}_{k}
- \frac{1}{\bar a^2} \vec \nabla^2 \bar h^{j}{}_{k} = 0
\ee
i.e. the propagation equations for a set of minimally
coupled scalars. The index structure 
of the perturbations $\bar h^{j}{}_{k}$
can be easily accounted for by going to the mode
expansion $\bar h^{j}{}_{k} = \bar \epsilon^{j}{}_{k}(t, \vec p)
\exp(i \vec p \cdot \vec x)$. The gauge conditions 
$\bar \epsilon^{j}{}_{j} = p_{j} \bar \epsilon^{j}{}_{k} = 0$
can be easily solved as follows. We orient the spatial reference
frame s.t. the $z$ axis is along the direction of propagation
of the wave, and then we find that two linearly independent 
polarization tensors are given by two Pauli matrices 
$\epsilon_{+} = \sigma_3$ and $\epsilon_{\times} = \sigma_1$.
Any other perturbation is their linear combination:
$\bar \epsilon^{j}{}_{k} = f_+ \epsilon_+ 
+ f_\times \epsilon_\times$,
where $f_k$, $k = (+,\times)$ are the mode functions. 
Since we will see later that the mixed index perturbations 
$\bar h^{\mu}{}_{\nu}$ are conformally invariant, and 
$f_k$ are modes of these degrees of freedom, $\bar f_k = f_k$
and from now on we will omit the bars from $f_k$'s. In a
different coordinate system, the basis polarizations are 
given by $\epsilon_k = R^{-1}(\vec p) \sigma_k R(\vec p)$,
where $R(\vec p)$ is the rotation matrix which orients 
$\vec p$ along the $z$ axis. Then, as we mentioned above,
the equations of motion (\ref{Epeexfin})
can be rewritten in terms of the mode functions
$f_+, f_\times$ as the Klein-Gordon 
equations for a set of minimally
coupled massless scalar fields, exactly as in General
Relativity:
\be
\label{KGEfree}
\bar \nabla^2 f_k = 0
\ee
This equation suggests very strongly that the metric perturbations
choose to propagate along geodesics 
in the Einstein conformal frame. We will
demonstrate that this is indeed true in the next section. 

Now we turn our attention to the description of 
the wave propagation from the point of view of the JBD frame.
We could proceed in precisely the same way as in the derivation
of the wave equation in the E frame. Starting with a fixed
JBD background, we add a perturbation to the metric as
$\hat g_{\mu\nu} = g_{\mu\nu} + h_{\mu\nu}$, and expand
the equations of motion (\ref{JBDeom}) to first order in
the fluctuation. Here, however, we don't have to repeat all
the steps of the derivation in the E frame, because most
of the details are the same, with barred quantities (E frame)
replaced with the unbarred ones. We will therefore outline
only the main points and establish a correspondence between
the two pictures. First, note that since the field redefinitions 
(\ref{conft}) imply that $h_{\mu\nu} = \bar h_{\mu\nu}/\chi$,
the mixed index perturbations are conformally 
invariant: $h^{\mu}{}_{\nu} = \bar h^{\mu}{}_{\nu}$. Next,
the transverse traceless conditions in the JBD
synchronous gauge are also conformally invariant: clearly,
the equality of the mixed index tensors implies that if one
is traceless, so is the other. Further, the constraints of the
JBD synchronous gauge on $h^{\mu}{}_{\nu}$ are also identical:
$h^{0}{}_{0}=h^{0}{}_{j}=0$. Finally, since 
\ba
\label{confcond}
\nabla_{\mu} h^{\mu}{}_{\nu} &=&
\bar \nabla_{\mu} \bar h^{\mu}{}_{\nu} 
+ \gamma^{\mu}_{\mu\rho} \bar h^{\rho}{}_{\nu}
- \gamma^{\rho}_{\mu\nu} \bar h^{\mu}{}_{\rho} \nonumber \\
&=& \bar \nabla_{\mu} \bar h^{\mu}{}_{\nu} 
- 2 \bar h^{0}{}_{\nu} \frac{\dot \chi}{\chi}
+ \bar h^{\mu}{}_{\mu} \frac{\partial_{\nu}\chi}{2\chi} =
\bar \nabla_{\mu} \bar h^{\mu}{}_{\nu}
\ea
the transversality
condition is also the same: $\nabla_{\mu} h^{\mu}{}_{\nu} = 0$.
Hence the perturbation of the Ricci tensor in the JBD frame
is of the same form as (\ref{Erich}), except that the quantities are
all unbarred. There however are additional source terms, since
now we have to take into account the contributions of the
second derivatives of the field $\chi$ as presented in
(\ref{JBDeom}). These terms will be important for 
showing that the geometrical optics limit of the
wave dynamics picks the E frame evolution. Using
the conformal correspondence we have established so far,
we can write down the equations of motion for the
perturbations in the JBD frame:
\ba
\label{JBDpert}
\nabla_{\lambda} \nabla_{\mu} h^{\lambda}{}_{\nu} +
\nabla_{\lambda} \nabla_{\nu} h^{\lambda}{}_{\mu} &=& 
\nabla^2 h_{\mu\nu} + \frac{\rho + p}{2} h_{\mu\nu} 
+ h_{\mu\nu} \frac{\nabla^2 \chi}{\chi}
\nonumber \\
&+&\frac{\nabla_\rho \chi}{\chi} \Bigl(
\nabla^{\rho} h_{\mu\nu} - \nabla_{\mu} h^{\rho}{}_{\nu}
- \nabla_{\nu} h^{\rho}{}_{\mu} \Bigr)
\ea
In the FRW background, after some straightforward 
algebra, we can rewrite these as: 
\be
\label{JBDpeex}
{h''}^{j}{}_{k} + 3 H {h'}^{j}{}_{k}
- \frac{1}{a^2} \vec \nabla^2 h^{j}{}_{k} + 
\frac{\chi'}{\chi} {h'}^{j}{}_{k}
+ (2 H' + 6 H^2 + \frac{\chi''+5H\chi'}
{\chi} + \frac{p - \rho}{2\chi}) 
h^{j}{}_{k} = 0
\ee
Again, the environment-induced mass term vanishes, now  
by way of the JBD frame FRW equations of motion (\ref{JBDefrw}).
The final set of the equations of motion for the
transverse traceless perturbations of the metric are
\be
\label{JBDpeexfin}
{h''}^{j}{}_{k} + 3 H {h'}^{j}{}_{k}
- \frac{1}{a^2} \vec \nabla^2 h^{j}{}_{k} + 
\frac{\chi'}{\chi} {h'}^{j}{}_{k} = 0
\ee
These equations contain the
term proportional to $\chi'/\chi$, since in the JBD frame
the tensor perturbations 
also couple to the scalar $\chi$. 
To see how this term arises in the form above, recall
that because (\ref{Epeexfin})
and (\ref{JBDpeexfin}) map into each other under 
conformal transformations (\ref{conft}), the $\chi'$-dependent
term must arise from the effect of the conformal map on the 
comoving time. Now, it is evident 
that the same polarization basis
as that used in the E frame analysis above
can be employed to represent an arbitrary perturbation.
In fact, we could just take any solution of (\ref{Epeexfin}), 
and conformally transform it to the JBD frame according
to (\ref{conft}), and it will be guaranteed to solve 
(\ref{JBDpeexfin}) as well. 
In terms of the mode functions, then, 
the JBD frame equation
of motion for gravity waves (\ref{JBDpeexfin}) can
be written as
\be
\label{JBDKG}
\nabla_{\mu}\Bigl(\chi \nabla^{\mu} f_k \Bigr)=0
\ee
where $f_k$ are the mode functions, with
$k\in \{+,\times\}$. In this case, the modes
propagate under the influence of an additional
conformal coupling to $\chi$. This equation
is conformal to (\ref{KGEfree}) under (\ref{conft}).
These two equations, (\ref{KGEfree}) and 
(\ref{JBDKG}), will be our starting point
of derivation of the geometrical optics approximation
in the next section.

\section{Singularity Revealed}

We are finally ready to show that smoothing the JBD frame is
not sufficient to remove the singularity from FRW cosmological
solutions we have discussed in section 2. As we have indicated 
in the introduction, this is because the
gravitons, when considered as probes of the geometry
in the classical limit, move along null geodesics of the
E frame, and not JBD. Since the E frame is manifestly
singular, the graviton worldlines are incomplete and so
they reintroduce the singularity's effects even in the
JBD frame. 

Let us first derive the geodesic equations and then consider
their implications. As a warm up, we will first find the
geodesics in the E frame, since there the mode equation
is very simple - it is just the Klein-Gordon equation for
a minimally coupled massless scalar field for each 
polarization mode of the graviton sector, $\bar \nabla^2 f_k =0$.
Now, the geometrical optics limit corresponds to setting 
$f_k = \exp(i S)$, and identifying the phase $S$ as the
action of the pointlike probe which 
replaces the wave packet \cite{ckmo}.
The field equation in terms of $S$ becomes 
$i \bar \nabla^2 S - (\bar \nabla S)^2 = 0$, and so,
assuming that $S$ is real and separating the real and
imaginary parts of the equation we find 
\be
\label{Egeom}
(\bar \nabla S)^2=0~~~~~~~~~~~~~~~\bar \nabla^2 S=0
\ee
Now, after S is integrated, it must
be a function of the space-time coordinates only,
since it is the phase of $f_k$. Hence, if we 
use $S=S(x^{\mu})$, the action must be representable
as a path integral along a geodesic which the wave packet
is following: $S = \int dx^\mu \bar \nabla_\mu S$. Let
us now introduce $V_{\mu} = \bar \nabla_\mu S$ and recall
that $dx^\mu = \dot x^{\mu} d \bar \lambda$ along a geodesic,
where $\bar \lambda$ is the affine 
parameter. Hence \cite{ckmo},
\be
\label{Egeom1}
S = \int V_\mu \dot x^\mu d \bar \lambda
\ee
The equations of motion for $S$ (\ref{Egeom}) 
become, in terms of the
field $V_\mu$, the following two constraints: 
$\bar g^{\mu\nu} V_\mu V_\nu = 0$, 
$\bar \nabla_\mu \bar g^{\mu\nu} V_\nu =0$.
The first constraint is local and hence can be easily 
enforced at the level of the particle action with the
help of a Lagrange multiplier. It just tells us that $V_\mu$
is a null vector. The second constraint is not local, and
in fact in the geometrical optics limit is always
a very small quantity compared to the first one, and is 
usually ignored. Here we will retain 
it, and use it to determine the
Lagrange multiplier. The constrained particle action is then 
$S=\int d\bar \lambda (V_\mu \dot x^{\mu} + \bar \eta 
\bar g^{\mu\nu} V_\mu V_\nu)$.
In the geometrical optics approximation, because of destructive
interference of waves, only those trajectories for which $S$
is extremized survive. Treating $V_{\mu}$ as an independent
variable, and varying $S$ with respect to it, we find
$\bar g^{\mu\nu} V_\nu = -\dot x^\mu /2 \bar \eta$. 
This simply means
that $V_{\mu}$ is tangent to the wave packets world-lines, 
which by $\bar g^{\mu\nu} V_{\mu} V_{\nu}=0$ must be null. 
The differential (second) constraint 
translates into the condition 
$\bar \nabla_\mu (\dot x^\mu /\bar \eta)=0$. Now, 
in the flat space limit, the covariant derivative 
would have become an ordinary derivative, and we would have
been able to use $\partial_\mu \dot x^\mu = 
d (\partial_\mu x^\mu)/d\bar \lambda = 0$ to assert that
$\dot x^\mu \bar \partial_\mu (1/\bar \eta)= 
d(1/\bar \eta)/d\bar \lambda = 0$, 
i.e. that $\bar \eta$ is a constant 
along trajectories. Then, with an appropriate normalization
($\bar \eta = -1/4$) we would have found $S=\int d \bar \lambda 
\dot x^\mu \dot x_\mu$ with $\dot x^\mu \dot x_\mu =0$,
i.e. just the standard action of a relativistic massless
particle. In the curved space, the tangent vector field
of a congruence of geodesics need not be divergenceless
in general, i.e. it need not be $\bar \nabla_\mu \dot x^\mu =0$.
However, since this quantity is the 
second constraint in (\ref{Egeom}),
because of destructive interference, 
its contribution to wave propagation
is negligible in the geometrical optics
approximation.
Moreover, if we go to the Riemann 
normal coordinates, defined at a point
${\bf x_0}$ by the condition 
$\bar \Gamma^{\mu}_{\nu\lambda}({\bf x_0}) = 0$,
we will indeed find that to the 
lowest order $\bar \nabla_\mu
\dot x^\mu$ vanishes. So, we will set  
$\bar \nabla_{\mu} \dot x^\mu =0$, 
as a part of the geometrical
optics approximation. This then tells us that 
$\dot x^\mu \bar \nabla_\mu (1/\bar \eta)= 
d(1/\bar \eta)/d\bar \lambda = 0$ - just as in the flat 
space limit, $\bar \eta$ is a constant along 
geodesics. Therefore, the particle action in the E frame
becomes, after again choosing $\bar \eta = -1/4$, \cite{ckmo}
\be
\label{Epart}
S = \int d\bar \lambda \bar g_{\mu\nu} 
\dot x^\mu \dot x^\nu ~~~~~~~~~~~~~
\bar g_{\mu\nu} \dot x^\mu \dot x^\nu = 0
\ee
which defines dynamics of the E frame massless
minimally coupled particles - i.e. the null geodesics.
The geodesic equations can be derived straightforwardly
by varying this action and imposing 
$\bar g_{\mu\nu} \dot x^\mu \dot x^\nu = 0$.
The result is 
\be
\label{Egeod}
\ddot x^\mu + \bar \Gamma^{\mu}_{\nu\lambda} 
\dot x^\nu \dot x^\lambda = 0  ~~~~~~~~~~~~~
\bar g_{\mu\nu} \dot x^\mu \dot x^\nu = 0
\ee
Hence as claimed, in the geometrical optics limit the
quanta of gravitational perturbations move along 
the E frame null geodesics, and not along trajectories
in the JBD frame.

We can now derive the same result using the JBD form
of the equation of motion, in order to check how the
dilaton force deforms the JBD graviton 
trajectories into the E frame null geodesics. To show
that the result is a conformal transform of 
(\ref{Egeod}), we will derive it from first
principles, rather than just apply the field
redefinition (\ref{conft}). Bearing in mind
that we are following the motion of the
same mode ($h^{\mu}{}_{\nu} = \bar h^{\mu}{}_{\nu}$), 
but in a different reference frame,
we again use $f_k = \exp(iS)$, but now
we substitute it in the JBD frame equation 
$\nabla_\mu (\chi \nabla^\mu f_k) = 0$ 
from Eq. (\ref{JBDKG}). When $S$ is
real, we again get two equations,
\be
\label{JBDgeom}
(\nabla S)^2=0~~~~~~~~~~~~~~~ 
\nabla_\mu (\chi \nabla^\mu S)=0
\ee
The first of these two equations is identical, up to an
overall factor of $\chi$, to the first equation in
(\ref{Egeom}). Thus it also must correspond to the null
condition, $g^{\mu\nu} V_\mu V_\nu = 0$ where we still have 
$V_\mu = \nabla_\mu S$, 
albeit in the JBD frame. The second equation
is a little bit more complicated than in (\ref{Egeom}).
To see the effect of the coupling $\chi$, let
us rederive the equations of motion for graviton
probes along the same lines we followed in the E frame
calculation above. As we have shown, S is 
a function of the space-time coordinates only, and when we 
use $S=S(x^{\mu})$, $S$ again is just the (same) integral 
$S = \int dx^\mu \nabla_\mu S$. 
This of course must be identical, since
$\nabla_{\mu} S = \partial_{\mu} S = \bar \nabla_{\mu} S$.
With $V_{\mu} = \nabla_\mu S$ and 
$dx^\mu = \dot x^{\mu}d \bar \lambda$ along a trajectory,
\be
\label{JBDgeom1}
S = \int V_\mu \dot x^\mu d \bar \lambda
\ee
The JBD frame equations of motion for $S$ then are 
$g^{\mu\nu} V_\mu V_\nu = 0$, 
$\nabla_\mu (\chi g^{\mu\nu} V_\nu)=0$.
Enforcing the first constraint at the 
level of the action with the
help of a Lagrange multiplier, we find 
$S=\int d\bar \lambda (V_\mu \dot x^{\mu} 
+ \eta g^{\mu\nu} V_\mu V_\nu)$.
Note that here we are contracting the indices with the
JBD frame metric $g_{\mu\nu}$, and so $\eta \ne \bar \eta$.
Varying this action with respect to $V_\mu$, we find
$g^{\mu\nu} V_\nu = -\dot x^\mu/2 \eta$, and so again, 
$V_{\mu}$ is a null vector tangent to the graviton
trajectories in the JBD frame. However, the 
differential constraint is then 
$\nabla_\mu (\chi \dot x^\mu/\eta)=0$.
We still implement $\nabla_\mu \dot x^\mu =0$ as a part 
of the geometrical optics
approximation, and find $\dot x^\mu 
\bar \nabla_\mu (\chi/\eta)= 
d(\chi/\eta)/d\bar \lambda = 0$, 
implying that $\chi/\eta = c$ 
is constant along each particle path. Choosing $c = -4$,
we can write the JBD frame particle action as 
\be
\label{JBDpart}
S = \int d\bar \lambda \frac{1}{\chi} g_{\mu\nu} 
\dot x^\mu \dot x^\nu ~~~~~~~~~~~~~
g_{\mu\nu} \dot x^\mu \dot x^{\nu} = 0
\ee
The particle trajectories are those paths which extremize
this action. By varying this action, and 
taking the null constraint into account, we find \cite{ckmo}
\be
\label{JBDgeod}
\ddot x^\mu + \Gamma^{\mu}_{\nu\lambda} 
\dot x^\nu \dot x^\lambda - 
\frac{\nabla_\nu \chi}{\chi} \dot x^\nu
\dot x^\mu = 0 ~~~~~~~~~~~~~
g_{\mu\nu} \dot x^\mu \dot x^\nu = 0
\ee
Hence in the JBD frame gravitons do not
move along geodesics, but rather along null trajectories
determined by the additional force proportional
to the $4$-gradient of the scalar field $\chi$. Of course,
this is what we have expected all along, since we
can see that in the JBD frame action (\ref{JBDact})
the $\chi R$ coupling implies that the JBD ``gravitons"
have $\chi$ field charge, and hence must couple to
$\chi$'s field strength. Yet, when we use the
conformal transformation (\ref{conft}) which 
transforms the JBD action (\ref{JBDact}) to the
E frame action (\ref{Eact}), rendering the
graviton kinetic term canonical, i.e. just
$\bar R$, we see that this same redefinition
removes completely the $\chi$-dependent force from
the equation (\ref{JBDgeod}). Under the
conformal transformation (\ref{conft}),
the connexion changes according to 
$\Gamma^{\mu}_{\nu\lambda} = \bar \Gamma^{\mu}_{\nu\lambda} 
+ (1/2 \chi) (\delta^{\mu}_{\nu} \nabla_\lambda \chi + 
\delta^{\mu}_{\lambda} \nabla_\nu \chi
- \bar g^{\mu\rho} \bar g_{\nu\lambda} \nabla_{\rho} \chi)$
and when contracted with $\dot{x^\nu} \dot{x^\lambda}$, the
difference precisely cancels the $\chi$-dependent
force in (\ref{JBDgeod}). Hence, the equations 
(\ref{Egeod}) and (\ref{JBDgeod}) are conformal images of
each other, and they imply that the gravity wave 
packets, to the lowest order, move along null geodesics 
of the Einstein frame metric $\bar g_{\mu\nu}$.

Having proven this, we conclude that the conformal removal
of the singularity was only partially successful. By
the construction of the theory as given in the action
(\ref{JBDact}), the matter fields ${\cal Y}$ coupled only
to the JBD frame metric $g_{\mu\nu}$. This metric was made 
smooth by the choice of the coupling function $\omega(\chi)$,
whose role was to push the E frame singularity to the asymptotic
timelike infinity of the geometry. Due to this effect,
the point-like matter probes were unable to ever reach the
singular region. Since they follow the JBD frame
geodesics, these geodesics are all complete. However,
in this case, geodesic completeness is insufficient to
conclude that the solution is nonsingular. 
The gravity waves couple to both the metric and the scalar
field in the JBD frame, and the net effect of these 
couplings is to deform the graviton trajectories back to
the E frame null geodesics. These worldlines are incomplete,
because the E frame metric has a singularity a finite
affine distance from any other place in the manifold.
As a result, we cannot arbitrarily extend the history
of one such Universe. Sooner or later,
we will reach the E frame singularity, where we will
have to deal with the problem of defining the proper
initial conditions for gravity waves. Because the graviton
sector there is ill-defined, we simply wouldn't be able to 
unambiguously set the initial conditions.
Gravity waves could then communicate the
presence of the singularity to the other degrees of freedom 
in the theory, at higher order. To see
this, note that because the matter fields couple to the metric,
they would also couple to the tensor perturbations at
higher order. Since to linear order the
tensor perturbations were the only degrees of freedom which 
detected the singularity, none of the other modes could be
adjusted to completely cancel the influence of the singularity
on the gravitons. Therefore the gravitons would render
the hidden singularity again visible to all the fields
in the theory. In particular, the relic gravitons from 
a very early era of the Universe that survived 
recombination and that comprise the present graviton 
background would still encode information about the 
maelstorm they came from.

\section{Conclusion}

In this article, we have shown that conformal
transformations alone cannot completely remove the initial
singularity from a cosmological solution in the scalar-tensor
models, even if we allow the scalar-gravity coupling strength
to depend on the scalar field. The matter sector in the models
we have considered consists only of modes whose stress-energy
tensor satisfies the strong energy condition. Clearly, if
the matter sources violate the SEC, then evading singularity
may yet be possible - but this cannot be accomplished by a simple
conformal transformation involving the JBD scalar. Such conformal
transformations merely hide the singularity from observers which
propagate along geodesics of the JBD frame metric, that can
be arranged to be smooth by adjusting the coupling $\omega(\phi)$.
However, once different observers are allowed, which couple
to both the JBD frame metric and scalar, because of the additional
scalar force they don't move along the JBD frame geodesics, but
along those of a different conformal frame. In particular, gravitons
move along the Einstein frame null geodesics, which are always
incomplete, because the E frame solutions are all singular. Hence
gravity waves see the singularity in all scalar-tensor models,
regardless of the specifics of the model in question. We can in fact
see that a similar property should hold in any 
nonminimal effective theory of gravity. The plain vanilla 
field redefinitions cannot 
remove singularities because they do not couple universally
to all the fields in the theory, and so there will always be a mode
which will discern the presence of the singularity in the
manifold. 

In string theory, however, this may merely mean that
the effective action approach must break 
down close to the singularity.
Since all the states in string theory
are comprised of strings (and D-branes, as we have 
seen recently in some of the
models), close to the singularity, gravity's pull on
low energy point-like degrees of freedom is so 
strong that they effectively
decompactify. Thus near the singularity 
instead of a gas of interacting highly 
energetic particles we find a gas 
of interacting highly energetic strings. 
But strings couple
naturally to the string frame metric, 
which therefore seems to be
the frame we must choose to study the 
effects of finite size of probes
(or higher order $\alpha'$ terms in 
the derivative expansion). However,
strings also couple to the dilaton, 
and we must consider its 
effect on dynamics too, since it 
represents the string coupling constant
and so controls the validity of the 
semiclassical approximation. 
Therefore in string theory, if we 
are to remove the
singularity, we must regulate both 
the metric and the dilaton, 
which is reminiscent of the situation
in the scalar-tensor models we 
have considered here.
This also implies that if we 
manage to cure the solution in one frame, 
as long as we take only 
reasonable field redefinitions 
(i.e. those which don't 
alter the global properties of the space-time) 
we find that the solution 
is free of singularities in any frame. 
This, of course, is consistent with the fact that
field redefinitions do not change physics,
but just alter the language we describe physics with.

Finding whether the singularities 
could be removed from
the effective theory still awaits. 
What we have learned to date 
seems to indicate that to regulate 
the singularities we must impose
severe alterations on the theory. Simple 
tweaks of the classical 
(or semi-classical) effective
action do not seem to work. 
To illustrate this point,
we may recall the graceful 
exit problem in the Pre-Big-Bang
scenario \cite{gv}. The idea, briefly, was to use the 
multiplicity of solutions which arise 
because of duality, and paste them together 
in such a way that the result is a smooth Universe
of an infinite life span and 
with a region of very large 
coupling {\it and} curvature, 
mimicking the Big Bang. The 
difficulty with this proposal 
is that the continuity of the
solutions and the equations of 
motion, in the effective 
potential approximation (i.e. 
the effective action truncated 
to second order in derivatives) 
makes the smooth matching
of the two branches with the 
proper asymptotic behavior
impossible unless very exotic conditions
develop at very high energies \cite{kmo}. 
The matching
is supposed to take place very
near the singularity - but in 
the semiclassical approximation,
the singularity does all the 
steering of the dynamics near it,
and hence the solutions cannot 
evade it. The branch changing
thus is not possible in the 
semiclassical limit, with the 
matter sources satisfying an 
even weaker version of the energy
conditions than in General Relativity. 
A necessary condition (but not sufficient)
for branch changing has been derived 
recently by R. Brustein and R.
Madden (in the last of Ref. \cite{kmo}),
who showed that unless 
the null energy condition 
(i.e. the requirement 
$\rho + p \ge 0$ for fluid sources)
is violated, the branch change 
cannot occur. This still
does not guarantee the evasion 
of the singularity via branch-changing, 
but it does tell us that unless we 
violate the null energy condition in some
way, in a region of high curvature, 
we cannot even hope to
avoid the singularity. Since the conventional
matter (point-like or string-like) degrees of
freedom generally do not violate the null energy condition,
this requires the presence of exotic matter sources 
(``string phase", \cite{vensp}) or nonperturbative phenomena 
(such as discussed by \cite{wind,art,kk})
to account for singularity smoothing. At this moment,
we know very little about this type of matter. The
recent rapid development of string theory however gives
hope that we may learn more in foreseeable future! 

\vspace{1cm}
{\bf Acknowledgements}

We would like to thank R. Brustein, 
H.S. Burton, R. Madden and R.C. Myers
for useful conversations. N.K. was supported 
in part by NSERC of Canada, and K.A.O. was supported in
part by  DOE grant DE--FG02--94ER--40823.

\end{document}